# Title: Anomalously Strong Near-Neighbor Attraction in Doped 1D Cuprate Chains

**Authors:** Zhuoyu Chen[1,2,3]†, Yao Wang[4]†, Slavko N. Rebec[1,2,3], Tao Jia[1,3,5], Makoto Hashimoto[6], Donghui Lu[6], Brian Moritz[1], Robert G. Moore[1,7], Thomas P. Devereaux[1,3,8]*, Zhi-Xun Shen[1,2,3,5]*

**Affiliations:**

[1]Stanford Institute for Materials and Energy Sciences, SLAC National Accelerator Laboratory, Menlo Park, CA 94025, USA.

[2]Department of Applied Physics, Stanford University, Stanford, CA 94305, USA.

[3]Geballe Laboratory for Advanced Materials, Stanford University, Stanford, CA 94305, USA.

[4]Department of Physics and Astronomy, Clemson University, Clemson, South Carolina 29631, USA

[5]Department of Physics, Stanford University, Stanford, CA 94305, USA.

[6]Stanford Synchrotron Radiation Lightsource, SLAC National Accelerator Laboratory, Menlo Park, California 94025, USA.

[7]Materials Science and Technology Division, Oak Ridge National Laboratory, Oak Ridge, Tennessee 37831, USA

[8]Department of Materials Science and Engineering, Stanford University, Stanford, California 94305, USA.

*Correspondence to: tpd@stanford.edu, zxshen@stanford.edu

†These authors contributed equally.

**Abstract:**

In the cuprates, one-dimensional chain compounds provide a unique opportunity to understand the microscopic physics due to the availability of reliable theories. However, progress has been limited by the inability to controllably dope these materials. Here, we report the synthesis and spectroscopic analysis of the one-dimensional cuprate $Ba_{2-x}Sr_xCuO_{3+\delta}$ over a wide range of hole doping. Our angle-resolved photoemission experiments reveal the doping evolution of the holon and spinon branches. We identify a prominent folding branch whose intensity fails to match predictions of the simple Hubbard model. An additional strong near-neighbor attraction, which may arise from coupling to phonons, quantitatively explains experiments for all accessible doping levels. Considering structural and quantum chemistry similarities among cuprates, this attraction will play a similarly crucial role in the high-$T_C$ superconducting counterparts

**One Sentence Summary:**

A quantitative comparison between theory and experiment reveals a key missing ingredient in the cuprates: a strong near-neighbor attraction.

**Main Text:**

The complexity of strong correlations in electronic materials like high-$T_C$ cuprates has been a central theme in condensed matter physics (*1*). However, so far, there remains no exact solution of many-body Hamiltonians in two dimensions (2D) and above, making a quantitative comparison between theory and experiment a challenge. An important, promising route to improved understanding may be through the analysis of one-dimensional (1D) systems: the reduced dimension enables exact solutions of microscopic Hamiltonians like the Hubbard model and low-energy field theory without bias. Such possibilities have been recognized in theory, but progress has been hindered by the lack of suitable experimental material systems.

In a 1D system, strong correlations manifest as so-called spin-charge separation (*2–7*). As schematized in Fig. 1A, a photoinduced hole fractionalizes into a "spinon" and a "holon", carrying spin and charge, respectively, giving rise to separate spectral branches with different propagation velocities. For half-filled $d^9$ systems, such spin-charge separation along with a Mott gap (Fig. 1B) has been confirmed by angle-resolved photoemission spectroscopy (ARPES) (*4–6*). However, the simplicity of the spectrum in the undoped case and the lack of information on interactions between doped carriers hinders a connection to underlying microscopic models. In particular, it remains undetermined whether fundamental Hamiltonians, like the Hubbard or *t-J* models contains all of the essential ingredients for general cuprate properties, including *d*-wave superconductivity in 2D. To answer this question, a doping dependent study of a 1D cuprate chain compound would be insightful. However, controllably doped 1D cuprate systems remain elusive after more than two decades of effort.

We achieved a wide range of hole doping in a corner-sharing cuprate chain by synthesizing $Ba_{2-x}Sr_xCuO_{3+\delta}$ (BSCO, whose lattice structure is shown in Fig. 1C). Thin films were grown with a recently developed ozone-reactive molecular beam epitaxy (MBE) system connected *in situ* to ARPES beamline 5-2 at the Stanford Synchrotron Radiation Lightsource (SSRL). Atomic-layer-by-layer growth in purified ozone was monitored in real-time via *in situ* reflective high energy electron diffraction (RHEED, Fig. S1B and C) (*8*). A mixture of isovalent Sr and Ba facilitates the tuning of in-plane lattice constants to mitigate mismatches with $SrTiO_3$ substrates (*9*). Orthorhombic BSCO on cubic $SrTiO_3$ forms twinned domains with different chain orientations (Fig. S1D and E). In different domains, chain directions are perpendicular to each other and parallel to the substrate surface (Fig. 1D). All ARPES data shown below were collected for temperatures lower than 20 K with 65 eV photons, from one single 2.5 unit-cell sample (equivalent to 5 $Ba_{1-x/2}Sr_{x/2}O$-CuO-$Ba_{1-x/2}Sr_{x/2}O$ layers) with nominal $x = 0.16$ measured by quartz crystal microbalance. Doping by interstitial oxygen was controlled via ozone and vacuum annealing series (*10*).

Figures 2A1–A6 present the ARPES Fermi surface maps with controlled hole doping from 9% up to 40%. The two perpendicular Fermi surfaces indicate simultaneous photoelectron detection from both sets of domains, as mentioned above. Within each set, the lack of transverse dispersion in the spectral distribution demonstrates the highly 1D nature of the chains. B1–B6 exhibit the energy-momentum spectra of cuts along the Brillouin zone edge. These cuts are located at a position where spectral weight from one set of domains is minimized, enabling analysis of the spectra from chains with a single orientation. Note that features at higher binding energies beyond 1.2 eV are compromised by non-bonding oxygen band intensities (*11*). C1–C6 show the corresponding second derivatives to highlight the quasiparticle features.

Comparison to cluster perturbation theory (CPT) simulations for the single-band Hubbard model (*12, 13*), displayed in Fig. 2D1-D6, allows for the following identification of spectral features: 1) the highest-intensity features correspond to main holon branches (blue dashed lines with "h" in B1 and C1) crossing the Fermi level $E_F$ at the Fermi momentum $k_F$; 2) spinon branches (red dashed lines with "s" in B1 and C1), which flatten near $k_\parallel = 0$ and merge into the main holon branches for higher momenta. These features agree with previous observations in undoped, parent materials (*4–6*). The persistence of spin-charge separation up to 40% doping agrees with predictions from the single-band Hubbard model, showing that correlation effects remain very strong in 1D, even for higher dopings (*14*). By fitting the doping-dependent spinon and holon binding energies at $k_\parallel = 0$ (Fig. S2, S3), we obtained Hubbard model parameters $U = 8t$ and $t = 0.6$ eV, with an estimate $J = 4t^2/U = 0.3$ eV consistent with earlier results (*6, 7*).

In contrast to the observations in undoped compounds (*4–6*), an additional feature at $|k|>|k_F|$, denoted by "hf" in Fig. 2B1 and C1, fades rapidly with doping, shown more directly in the momentum distribution curves (MDCs) for 9%, 14%, and 33% doping (see sub-dominant peaks in Fig. 3A, B, and C, respectively). As illustrated in Fig. 3D and in comparison with theory (*14–16*), the holon dispersion inherited from the undoped cuprate splits into two branches at $k_F$ for doped systems: one folding directly at $E_F$ and $k_F$ represented in green (hereafter the holon folding (hf) branch), and the other crossing $E_F$ and bending back at a larger momentum $2\pi - 3k_F$ (hereafter the $3k_F$ branch) as an extension of the main holon branch. Note that the term "folding" here is simply a description of the spectral feature rather than suggesting existence of translational symmetry breaking. The Lorentzian fitting of our MDCs yields the peak positions plotted in Fig. 3E and confirms the observed prominent "hf" spectral feature being the holon folding branch (green). The presence of the hf branch agrees with predictions using the single-band Hubbard (*14–16*) or *t-J* models (*17, 18*), previously attributed to a holon-holon interaction mediated by the spin superexchange and intrinsic in these models. This interaction occurs in the lowest-order $t/U$ expansion as an attractive form with the magnitude $\sim J/4 \sim 0.1t$ [see Supplementary Materials (SM)].

Despite a consistency with the basic dispersion, we find that the simple Hubbard model is fundamentally deficient in accurately addressing additional spectral features. As evident in the MDCs shown in Fig. 4A, a prominent spectral weight of the hf branch compared to the $3k_F$ branch (below the detection limit) in experiment is missing in the Hubbard model where the hf feature is barely visible. This implies that realistic holon-holon attractive interaction is not adequately described by the spin superexchange. To reproduce the experimental spectra, we find that a sizeable *attractive* near-neighbor Coulomb interaction $V$ must be introduced to the Hubbard Hamiltonian (see SM). Fig. 4A displays the MDCs extracted from simulations with different $V$ values from positive (repulsive) to negative (attractive): the near-neighbor *attractive* Coulomb interaction enhances the spectral weight of the hf branch, while suppressing the $3k_F$ branch. We conclude that a strong holon-holon attraction $V \approx -1.0t$ matches well with the experimental spectra, and explains why the $3k_F$ branch is not found. Note that in principle such attraction in 1D primarily affects the charge degree of freedom, thus the chemical potential is pinned by the spinon dispersions, unlike 2D cases where a quasiparticle gap is usually expected.

As mentioned earlier, the hf spectral feature fades rapidly with increasing doping, as summarized in Fig. 4B, perhaps implying a doping dependent holon-holon attraction. Comparing both the spectral intensity and the doping level at which it vanishes (Fig. S4, S5), we find the best agreement with a doping-independent $V$ between $-1.2t$ and $-0.8t$. To quantify this strong doping dependence, we extract the relative hf peak intensities with respect to the main holon-peak intensities based on Lorentzian fittings for experimental MDCs (circles in Fig. 4C); for comparison, we perform the same fitting on the simulated MDCs with $V = -1.0t$ (squares in Fig. 4C). The doping dependence of the hf peak intensity is captured for a fixed $V$, whose influence is renormalized by doping, such that this intensity gradually decreases and eventually disappears at ~ 30% hole-doping. In other words, the doping dependence is intrinsic in the calculation and does not require parameter tuning. The Coulomb interaction between electrons should always be repulsive, so the only probable origin of this effective attraction, beyond spin superexchange already accounted for in the Hubbard model, would be coupling to some bosonic excitations. While multiple bosonic modes may play a role in cuprates, here, we postulate that the attractive $V$ is mediated by phonons, considering the evidence of electron-phonon coupling in a variety of cuprates (*19–28*).

We emphasize that this additional near-neighbor attraction is an order of magnitude stronger than the inherent attraction in the Hubbard model mediated by spin superexchange (given by $-J/4$ in the limit of $t/U \ll 1$), when considering all interactions on the same footing. This is demonstrated in Fig. 4D, in which a baseline of zero net attraction is found by using the simulated folding peak relative intensity as a phenomenological parameter and back-extrapolating it as a function of $V$. In the case of only the Hubbard model ($V = 0$) the effective attraction is $\sim 0.1t$ above the baseline (consistent with the $J/4$ estimation), while the experiment case is at least $\sim 1t$ above the baseline, an order of magnitude larger. This difference implies that the single-band Hubbard model misses a sizable, attractive interaction between neighboring holes. Its magnitude is anomalously strong, and should be understood in the context of canonical values of electron-phonon coupling (*29, 30*).

The importance of this strong near-neighbor attraction cannot be overemphasized. Considering the structural similarities amongst cuprates, our Hamiltonian with near-neighbor attraction should be applicable to $CuO_2$ planes and favors neighboring electron pairs. Given that the Hubbard model with strong on-site repulsive $U$ and strong near-neighbor attractive $V$ explains the essential physics in 1D – charge, spin, charge-charge attraction and its doping dependence, it is likely that such an augmented Hubbard model provides a holistic picture for all cuprates, including robust *d*-wave superconductivity that has been elusive in all 2D Hubbard-like models.

**Acknowledgments:**

We thank S.A. Kivelson, D.H. Lee, and D. Orgad for stimulating discussions. ARPES experiments were performed at Beamline 5-2, Stanford Synchrotron Radiation Lightsource, SLAC National Accelerator Laboratory. **Funding:** This work is supported by the U.S. Department of Energy, Office of Science, Office of Basic Energy Sciences, Materials Sciences and Engineering Division, under Contract DE-AC02-76SF00515. This research used resources of the National Energy Research Scientific Computing Center (NERSC), a U.S. Department of Energy Office of Science User Facility operated under Contract No. DE-AC02-05CH11231. Research sponsored by the Laboratory Directed Research and Development Program of Oak Ridge National Laboratory, managed by UT-Battelle, LLC, for the U. S. Department of Energy. **Author contributions:** Z.C. conceived the experiment, grew the films, performed ARPES measurements, and analyzed data. Y.W. performed theoretical and numerical calculations. S.N.R. assisted in film growth and ARPES measurements. T.J. assisted in film growth. M.H. and D.L. developed the ARPES setup and assisted in ARPES measurements. B.M. assisted in calculations and data interpretations. R.G.M. developed the MBE setup and assisted in film growth. Z.C., Y.W., T.P.D., and Z.X.S. interpreted the data and wrote the manuscript with input from all authors. T.P.D. and Z.X.S. supervised the project on theory and experiment aspects, respectively. **Competing interests:** Authors declare no competing interests. **Data and materials availability:** The data that support the findings of this study are available from the corresponding author upon reasonable request.

**Supplementary Materials:**

Materials and Methods

Supplementary Text

Figures S1-S7

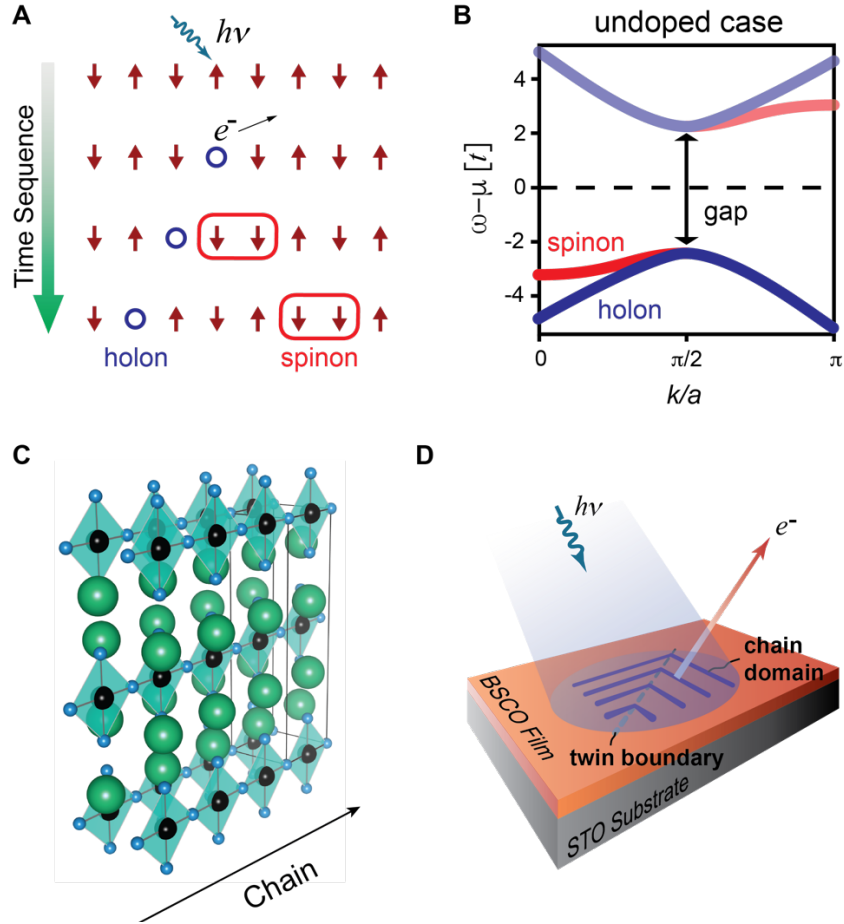

**Fig. 1 Strongly correlated antiferromagnetic (AFM) cuprate chain.**
(**A**) Schematic of an example photoemission process in a 1D AFM chain. Red arrows, blue circles, and red rounded rectangles correspond to spin-full electrons, holons, and spinons, respectively. The green arrow represents the arrow of time. (**B**) Schematic diagram of the spectral dispersion relation for an undoped chain. Blue and red solid lines correspond to the holon and spinon branches, respectively. (**C**) The lattice structure of $Ba_{2-x}Sr_xCuO_{3+\delta}$. Green, blue, and black balls correspond to Ba/Sr, O, and Cu atoms, respectively. (**D**) Schematic of sample setup for photoemission. The BSCO film on the $SrTiO_3$ (STO) substrate is twined with chain domains smaller than the beam spot.

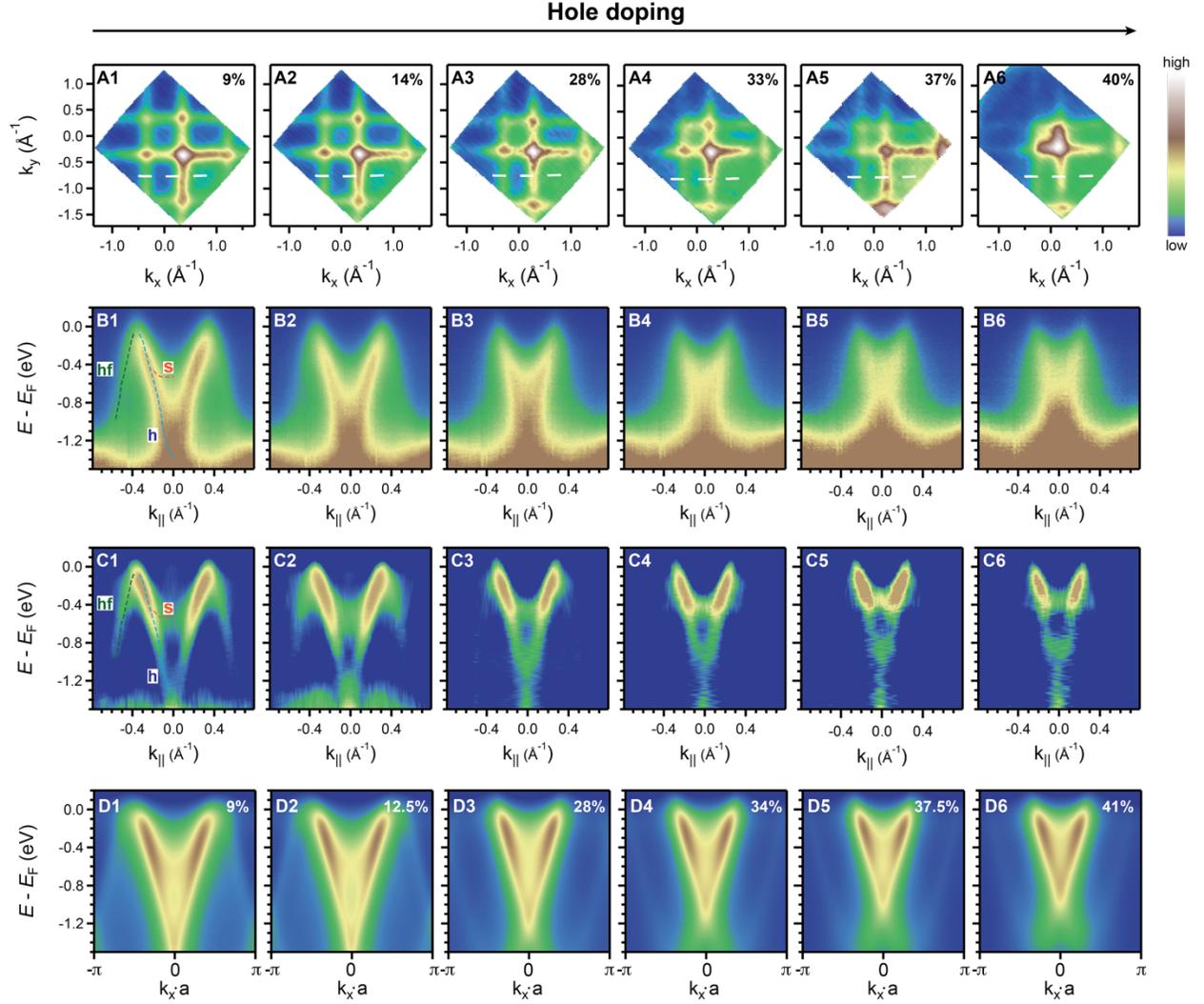

**Fig. 2 Hole doping series of ARPES spectra.**
(**A1-A6**) Fermi surface maps with different hole doping levels. (**B1-B6**). Measured ARPES spectra taken along the cuts shown with white dashed lines in (A1-A6). (**C1-C6**) Second derivatives with sign-reversed for comparison. Dashed blue, red, and green lines in (B1 and C1) indicate the location of the holon ("h"), spinon ("s"), and holon folding ("hf") features. (**D1-D6**). Spectral function simulation results based on 16-site cluster perturbation theory (CPT) with varied corresponding hole doping levels.

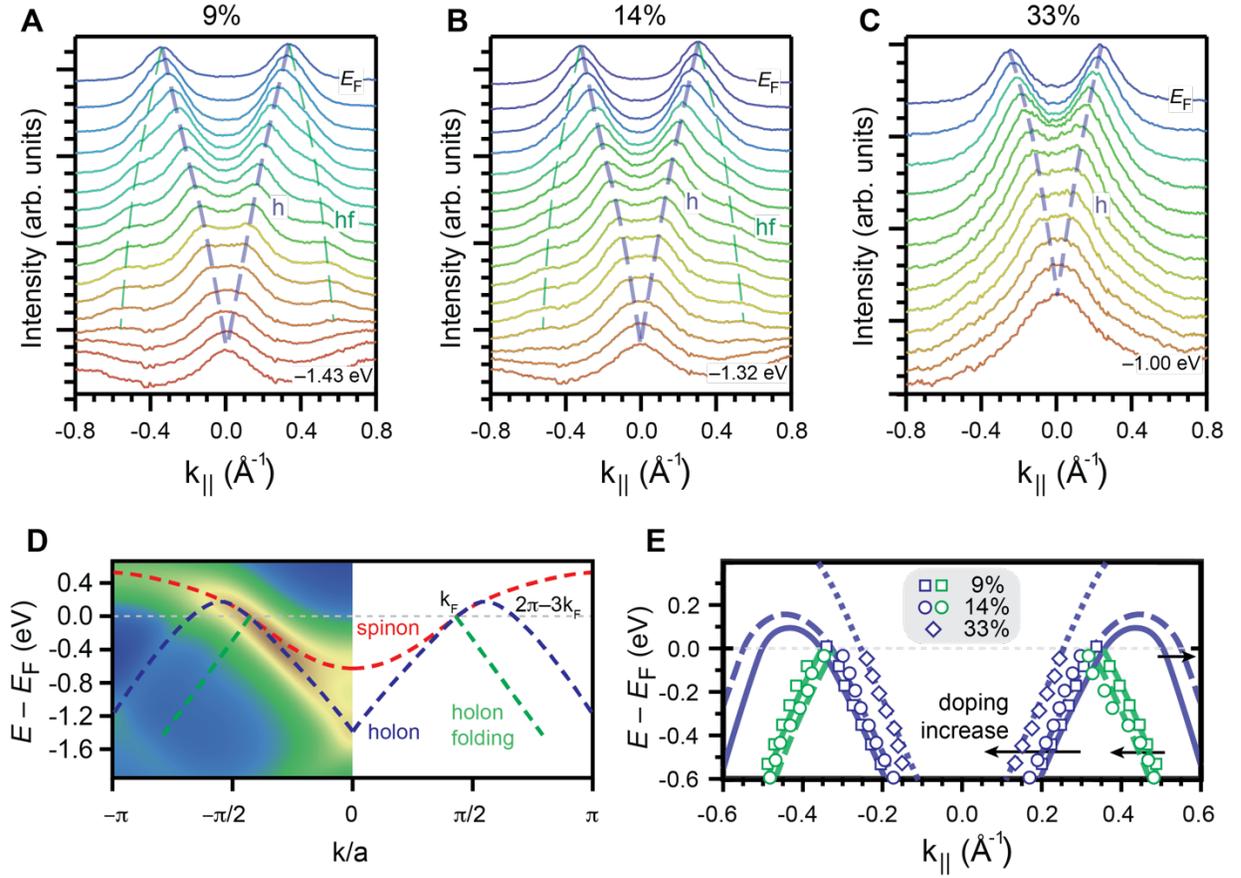

**Fig. 3 Identification of the holon folding.**
(**A-C**) MDCs for varied energies at 9%, 14%, and 33% hole doping. Blue and green dashed lines are guides to the eye for the holon branch (h), and the holon folding branch (hf). (**D**) CPT simulation based on single-band Hubbard model without Fermi-Dirac suppression above $E_F$ for the 12.5% doping case, shown in color scale plot. Dashed lines with different color correspond to schematic traces of the different branches. Note that the holon folding feature is relatively weak in the CPT simulation of pure Hubbard model without additional terms. (**E**) Peak positions extracted from MDCs for different dopings (squares: 9%, circles: 14%, diamonds: 33%) overlaid on schematic lines of single-band Hubbard prediction, where solid, dashed, and dotted lines correspond to increasing doping. Blue and green colors correspond to the major holon (h) and the holon folding (hf) branches, respectively. Black arrows indicate the direction of the shift for different features with increasing dopings.

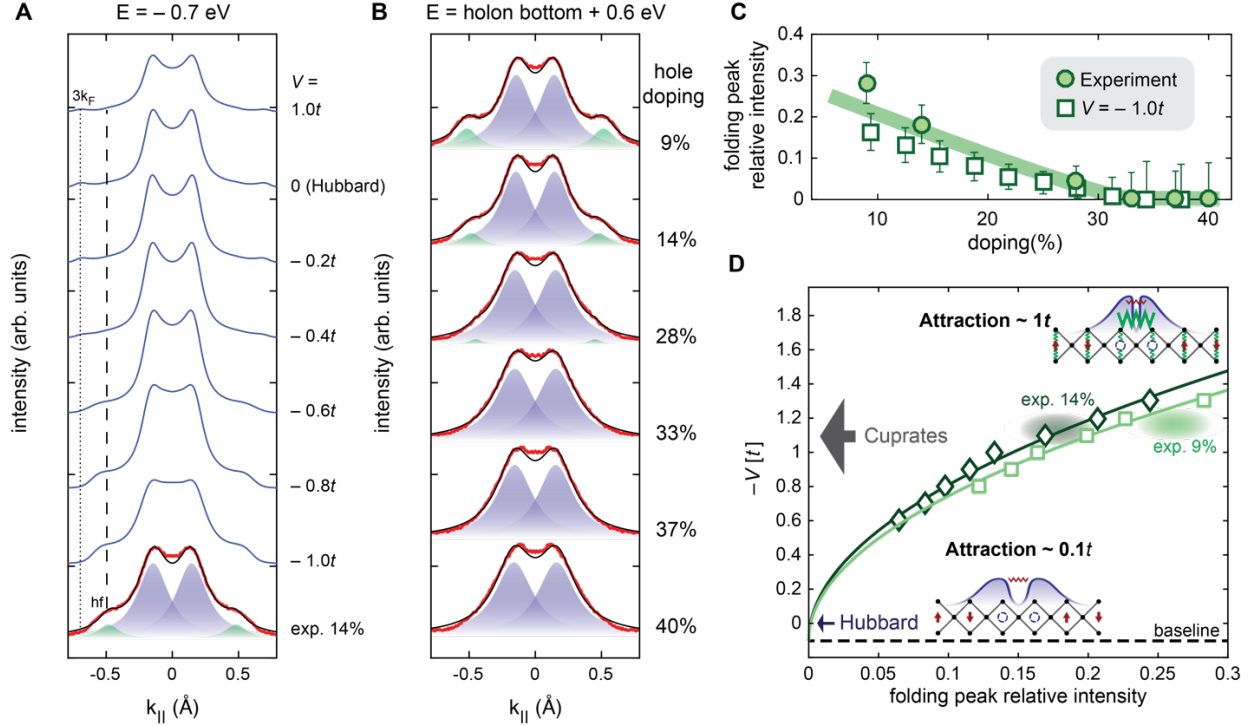

**Fig. 4 Strong holon folding indicating attractive interaction beyond Hubbard model.**
(**A**) CPT simulation incorporating a near-neighbor interaction term $V$ (blue curves), compared with experimental MDC at $-0.7$ eV below $E_F$ for 14% doping (red curve). Black curve is a fit utilizing two Lorentzian peaks shown in blue (major) and green (hf). Vertical dashed and dotted lines mark the locations of the hf and the $3k_F$ branch peaks, respectively. Curves are shifted and scaled vertically for comparison. (**B**) Doping-dependent experimental MDCs at energies approximately 0.6 eV above the holon bottom, such that the major branch peaks have similar momenta for comparison. (**C**) Folding peak intensity relative to the major peak intensity as a function of doping (circles: experiment; squares: simulations for $V = -1.0t$). The green line is a guide to the eye. (**D**) $-V$ as a function of CPT simulated folding peak relative intensity $f$ as a phenomenological measure of the net near-neighbor attraction (squares: 9%, diamonds: 12.5%). The solid green curves are a fit to the data using $f = a\,(V - V_0)^2$, where $a$ is a doping-dependent prefactor, and $V_0$ represents the compensation against the effective attraction in the Hubbard model, $\sim (0.10 \pm 0.05)t$ given by the fitting. The grey shaded areas correspond to experimental points (and errorbars) of 9% and 14% dopings. Insets: schematic of holon attractions for the pure Hubbard model (lower) and the mediation by bosons (upper). Besides same notions used in Fig.1A, green springs, blue shades, red, and green zigzags represent electron-phonon coupling, holon distributions, AFM-, and boson-mediated attractions, respectively.

# Supplementary Materials for

# Anomalously Strong Near-Neighbor Attraction in Doped 1D Cuprate Chains

Zhuoyu Chen[1,2,3†], Yao Wang[4†], Slavko N. Rebec[1,2,3], Tao Jia[1,3,5], Makoto Hashimoto[6], Donghui Lu[6], Brian Moritz[3], Robert G. Moore[3], Thomas P. Devereaux[1,3,7]*, Zhi-Xun Shen[1,2,3,5]*

†These authors contributed equally.
*Correspondence to: tpd@stanford.edu, zxshen@stanford.edu

**This PDF file includes:**

    Materials and Methods
    Supplementary Text
    Figs. S1 to S7

# Materials and Methods

## Thin Film Synthesis and ARPES Measurements

The synthesis is done with molecular beam epitaxy (MBE). The SrTiO$_3$ (STO) substrates are 0.05% Nb doped, mounted on inconel sample holders with silver conductive paste. We grow the Ba$_{2-x}$Sr$_x$CuO$_{3+\delta}$ (BSCO) films, at 635 °C, as measured by pyrometer with emissivity 0.7, with continuous purified ozone supply and the gas nozzle directly pointing at the sample, creating a $1 \times 10^{-5}$ Torr background pressure. Before the BSCO film growth, a thin buffer layer of either La:STO or LaTiO$_3$ is usually grown in an oxygen environment to ensure the TiO$_2$ termination and the atomic flatness. The sample shown in the main text is with 5 unit-cell LaTiO$_3$ buffer. The growth is performed atomic-layer-by-layer using a shuttered approach via *in situ* real-time reflective high energy electron diffraction (RHEED) monitoring, as shown in Fig. S1B. The doping is controlled via the temperature at which ozone supply is shut off. In particular, for the series of dopings shown in the main text, we grew the sample in ozone and cooled down in ozone until the sample temperature reached 100 °C, allowing us to achieve the highest 40% hole doping. After growth, the film is transferred *in situ* to the ARPES end station of the Stanford Synchrotron Radiation Lightsource beamline 5-2. After ARPES data collection was finished for this doping, we *in situ* transferred the sample back into the MBE chamber and annealed the sample in ultrahigh vacuum (UHV) at 180 ºC for about 1 hour to obtain the 37% doping. Then we *in situ* transferred the sample to the ARPES chamber again for measurement. This procedure was repeated over and over with increasing UHV annealing temperatures. Specifically, 270 °C, 350 °C, 430 °C, and 500 °C UHV annealing temperatures correspond to 33%, 28%, 14%, and 9% hole doping, respectively. The base pressure in the oxide growth chamber is about $1 \times 10^{-9}$ Torr. The base pressure in the ARPES chamber is lower than $4 \times 10^{-11}$ Torr. All ARPES data were collected for temperatures lower than 20 K with 65 eV photons and linear horizontal polarization with 10 eV pass energy. Equipment energy resolution is better than 40 meV. The angular resolution is better than 0.1°. In Fig. 2B1-B6 in the main text, a background reduction is done by subtracting halved-intensity energy distribution curve (EDC) of the Brillouin zone boundary (i.e. $k_\parallel \sim 0.78$ Å$^{-1}$). In Fig. 2C1-C6, second derivatives of EDCs and MDCs are summed to highlight spectra traces.

## Structural Characterizations

RHEED patterns of the BSCO film shown in the main text on both [100] and [110] directions (Fig. S1C, E) exhibit no sign of surface reconstruction. Due to the large beam spot, the RHEED detects both chain domains on the film simultaneously, thus it measures both in-plane orthorhomic lattice constants along the [100] direction. As shown in Fig. S1C, no splitting of higher order streaks (e.g. the (02) peaks) is seen, indicating identical in-plane lattice constants, most likely due to strain from the substrate. By comparing RHEED peaks separation before and after growth (Fig. S1D, E, [110] direction is preferred because it has larger seperation for lower error), we found the film top layer lattice constant is identical to the titanate substrate. Therefore, the in-plane lattice constants are $a = b = 3.9$ Å. Alternatively, in-plane lattice constants can be measured through Brillouin zone sizes in ARPES Fermi surface maps (Fig. 2A). The extracted value is independent of doping within error bars (estimated by ±1 pixel) and matches the value of SrTiO$_3$, as shown in Fig. S1F. For *ex situ* x-ray diffraction measurements (Fig. S1G, H) of thicker films, a Se capping layer is needed to protect the films from the atmosphere. Separate peaks in the reciprocal space map indicate different in-plane lattice constants for the orthorhombic domians in the thick films.

## Chemical Potential and Band Shifts

Fermi level was determined from the sample density of states (DOS) assuming Luttinger liquid (*3*). Based on Luttinger liquid theory, the DOS near the Fermi level obeys a power law dependence, i.e. DOS $\sim |E-E_F|^\alpha$. By fitting the angle integrated EDC near the edge where spectral intensity disappear, we extract the Fermi level for different dopings. The shift of $E_F$ is consistent with the change of the binding energy of the O 2p states, as shown in Figure S2.

# Supplementary Text

## Hubbard Model and Extended Hubbard Model

The Hamiltonian of the one-dimensional single-band Hubbard model is given by

$$\mathcal{H}_{\text{Hubbard}} = -t \sum_{i,\sigma} \left( c^\dagger_{i+1\sigma} c_{i\sigma} + h.c. \right) + U \sum_i n_{i\uparrow} n_{i\downarrow},$$

where $c^\dagger_{i\sigma}$ ($c_{i\sigma}$) and $n_{i\sigma}$ denote the creation (annihilation) and density operators at site $i$ of spin $\sigma$, respectively; $U$ denotes the on-site Coulomb interaction; $t$ denotes the nearest-neighbor electron hopping. The specific values of $U$ and $t$ are determined by comparing spectral functions with experiments (see Fig. S3).

In a 1D Hubbard model at half-filling, the photohole fractionalizes into a spinon and a holon degree of freedom, leading to the so-call spin-charge separation. In this case, the dynamics of spinon are governed by the super-exchange process whose energy scale is set by $J = 4t^2/U$, while the dynamics of a holon are governed by the hopping $t$. With different velocities, the spinon and holon give distinct spectral features in the single-particle spectral function (*4-6*). With doping, the low-energy physics of the Hubbard model can be captured by the *t-J* model, through a *t/U* expansion

$$\mathcal{H}_{t-J} = -t \sum_{i,\sigma} \left( c^\dagger_{i+1\sigma} c_{i\sigma} + h.c. \right) + J \sum_i \left[ \mathbf{S}_i \cdot \mathbf{S}_{j+1} - \frac{n_i n_{i+1}}{4} \right],$$

Here one immediately sees that the AFM in the Hubbard model naturally induces a – *J*/4 neighboring holon-holon attraction.

There is an additional effective interaction originated from the $\mathbf{S}_i \cdot \mathbf{S}_{j+1}$ term. In a simple Ising picture (with only the $S^z_i \cdot S^z_{j+1}$ type of interaction), the creation of a hole breaks two AFM spin bonds on its neighbors, which increases the energy by *J*. The proximity of two holes reduces the number of broken spin bonds by one, therefore lowers the energy by *J/2*. However, this effect is compensated by the three-site term, which is missing in the Hubbard to *t-J* model projection: the proximity of two holes blocks the three-site hopping of each, raising the energy by *J/2*. Furthermore, both effects rely on the spin order and are weakened by the spin fluctuations $S^+_i \cdot S^-_{j+1}$ and even more by the doping-induced scrambling of the spin correlations. Therefore, considering the cancellation and weakening, we ignore the contributions from the $\mathbf{S}_i \cdot \mathbf{S}_{j+1}$ term and the three-site term, estimating the effective, attractive near-neighbor interaction in the Hubbard as *J*/4.

To explain the intensive "hf" feature observed in experiments, we found it necessary to introduce an additional attractive interaction to augment the Hubbard model, as presented in Fig. 4 of the main text. Generally, we consider a Hamiltonian including longer-range density-density interactions as

$$\mathcal{H} = \mathcal{H}_{\text{Hubbard}} + V \sum_{\langle i,j \rangle} \sum_{\sigma,\sigma'} n_{i\sigma} n_{j\sigma'},$$

In the calculations presented in Fig. 4 of the main text, we truncate the interaction *V* at just the nearest neighbors to reflect the simplest ingredient of the model.

We remark that such a term can occur from longer-range Coulomb interactions which are strictly repulsive or from electron-phonon interactions where the phonons are integrated out. Most generally, the latter can produce an effective attraction interaction in many cases. An order-of-magnitude estimate of the effective interaction can be given by assuming $V \sim g^2/\omega$, where $g$ is the electron-phonon coupling strength and ω is the phonon frequency. Taking the phonon energy ω as 50meV, and the extracted effective interaction $V = -t$ (Fig.4 in the main text) leads to an estimate for the coupling of $g \sim 175$meV, which is consistent with the estimate from resonant inelastic x-ray scattering measurements on another 1D cuprate system (*27*). A more quantitative estimate of this e-ph coupling requires more rigorous theoretical calculations, taking into account the phonon retardation, the many-body nature of electrons, and their interplay. Likely due to the deconfined nature of the single-particle excitations in 1D, a phonon "kink" is not observed in our experiment.

**Spectral Function and Cluster Perturbation Theory**

To compare with ARPES experiments, we calculate the single-particle spectral function defined as:

$$A(k,\omega) = -\frac{1}{\pi}\text{Im}\sum_{\sigma}\langle G|c_{k\sigma}^{\dagger}\frac{1}{\omega+\mathcal{H}-E_G+i\delta}c_{k\sigma}|G\rangle,$$

where $c_{k\sigma} = \sum_j c_{j\sigma}e^{-ikj}/\sqrt{N}$ denotes the electron annihilation operator in momentum space, $|G\rangle$ is the ground state with energy $E_G$, and $N$ is the number of sites. Due to the relatively large broadening observed in experiments, we choose a Lorentzian energy broadening of $\delta = 0.3t$ to account for an intrinsic lifetime and an *ad hoc* Gaussian broadening of $0.2t$ for the overall spectral function to account for the extrinsic resolution.

In the main text, we employed the cluster perturbation theory (CPT) to estimate $A(k,\omega)$ due to its accessibility to continuous momenta (*12,13*), crucial for the comparison with experimental MDCs. CPT is constructed based on the fact that the electronic correlation ($U$) has a much large energy scale compared with the electron hopping $t$. Thus, the correlation effect is relatively local and can be captured by a small cluster. Therefore, one can divide the infinite plane into clusters and treat the hopping terms ($t$) on the boundary as a perturbation. Specifically, the Hamiltonian can be split into $\mathcal{H} = \mathcal{H}_c + \mathcal{H}_{\text{int}}$. Here $\mathcal{H}_c$ contains the (open-boundary) intra-cluster operators, while $\mathcal{H}_{\text{int}}$ contains the operators with inter-cluster indices (hopping terms for the Hubbard model). Then one can use exact diagonalization to exactly solve the Green's function $G_c(\omega)$ associated with the intra-cluster Hamiltonian $\mathcal{H}_c$. Then following the above perturbation, the CPT method estimates the spectral function $A(k,\omega)$ of the infinite chain by treating $\mathcal{H}_{\text{int}}$ perturbatively, giving

$$A(k,\omega)_{\text{CPT}} = -\frac{1}{\pi N}\text{Im}\sum_{a,b}\left[\frac{G_c(\omega)}{1-\sum_R e^{ikR}\mathcal{H}_{\text{int}}G_c(\omega)}\right]_{ab}e^{ik\cdot(r_a-r_b)},$$

where the indices in $\mathcal{H}_{\text{int}}$ are projected to the intra-cluster coordinates, and $R$ denotes the shifted unit vector for these boundary terms. When the boundary length is much smaller than the system size, such as in the case of 1D, $A(k,\omega)_{\text{CPT}}$ gives a good estimation of the $A(k,\omega)$. In the calculation of the main text, we employed a 16-site chain as the cluster for CPT. The comparison between calculated and experimental MDCs is shown in Fig. S5, S6, and Fig. 4 in the main text. The CPT calculations using different model parameters lead to the determination of $U$ and $t$ values in Fig. S3.

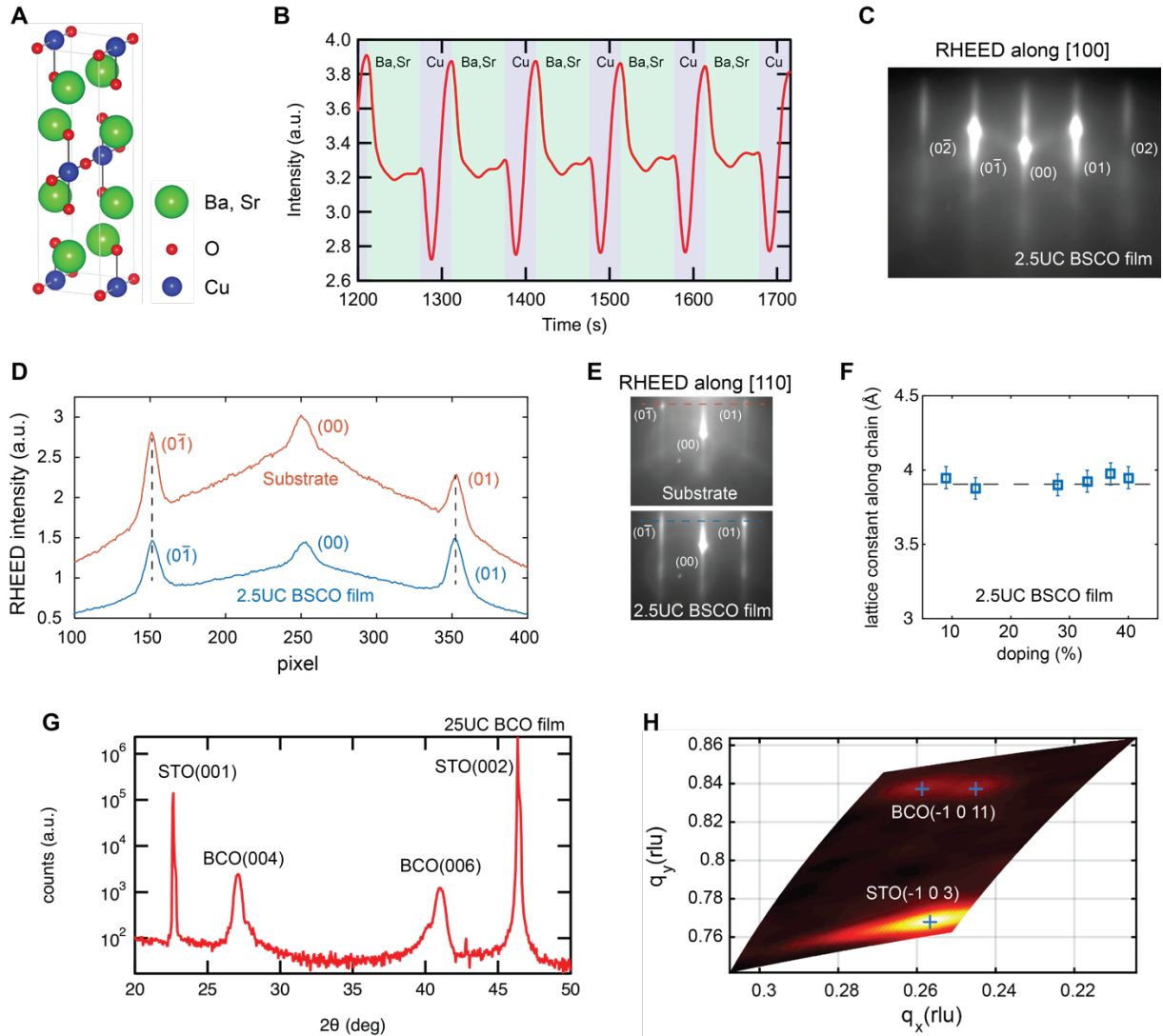

**Fig. S1.**
(**A**) Schematic structure of a unit cell of the $Ba_{2-x}Sr_xCuO_3$ compound. (**B**) Example reflective high energy electron diffraction (RHEED) intensity oscillations of the central specular spot as a function of time. Ba,Sr shutters and Cu shutter are opened alternatively as shown with a background of different colors. Within each opening period of the Ba,Sr shutters, two oscillations can be seen, indicating the growth of two $Ba_{1-x/2}Sr_{x/2}O$ atomic layers. (**C**) RHEED pattern along [100] direction after the growth of the ultrathin 2.5 unit-cell (UC) $Ba_{2-x}Sr_xCuO_{3+\delta}$ (BSCO) film, from which ARPES data shown in this manuscript are collected. No surface reconstruction is observed. (**D**) RHEED peaks along cuts shown with red and blue dashed lines in (E) for the titanate substrate and the 2.5 UC film, respectively. The identical spacing between the two (01) peaks indicate the film in-plane lattice constant is the same as the substrate, i.e. $a = b = 3.9$ Å. (**E**) RHEED patterns along [110] direction on substrate and the film. (**F**) Lattice constant along the chain extracted from ARPES Fermi surface maps in Figure 2A. Error bars are estimated based on ±1 pixel. Horizontal black dashed line indicates lattice constant of $SrTiO_3$. (**G**) X-ray diffraction along c-axis of a 25 UC $Ba_2CuO_{3+\delta}$ film grown on $SrTiO_3$ substrate, showing peaks corresponds to the 213 structure. (**H**) X-ray diffraction reciprocal space mapping of the same $Ba_2CuO_{3+\delta}$ film in (D), showing orthorhombic structure. Lattice constants obtained from this film are $a = 4.0$ Å, $b = 3.8$ Å, $c = 13.1$ Å.

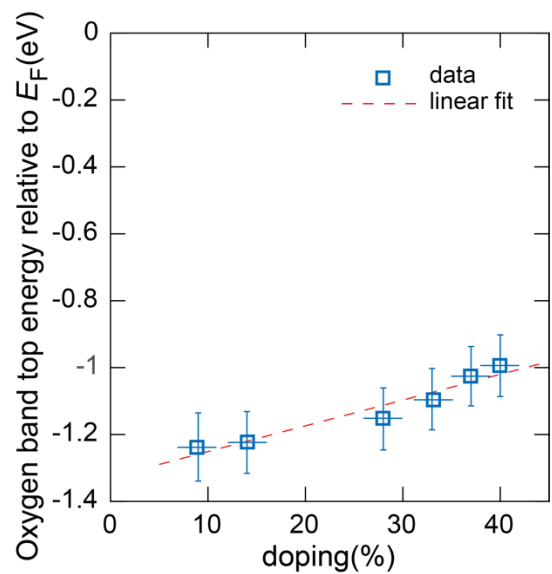

**Fig. S2.**
Doping dependence of the oxygen band top binding energy. Data (squares) are obtained by Gaussian peak fitting of first derivatives of the EDCs at the Brillouin zone boundary. Full width of vertical error bars corresponds to standard deviation of the fitted Gaussian peak. Red line is a linear fit to the data.

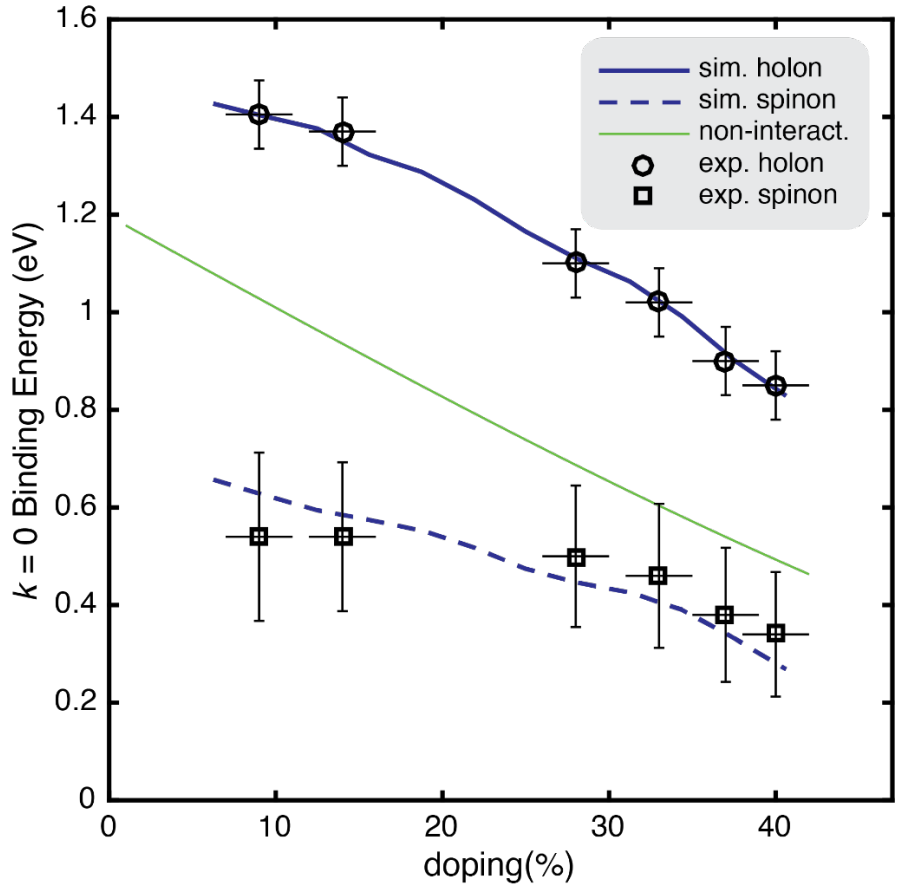

**Fig. S3.**
Holon and spinon binding energies at $k = 0$ as a function of doping. Open circles and squares are experimental data for holons and spinons, respectively. Blue solid and dashed lines are corresponding to the best fit of $U$ and $t$ to the experiment based on CPT simulations, giving $U = 8t$, and $t = 0.6$ eV. The green line represents the case without interaction, showing no splitting of spinon and holon branches.

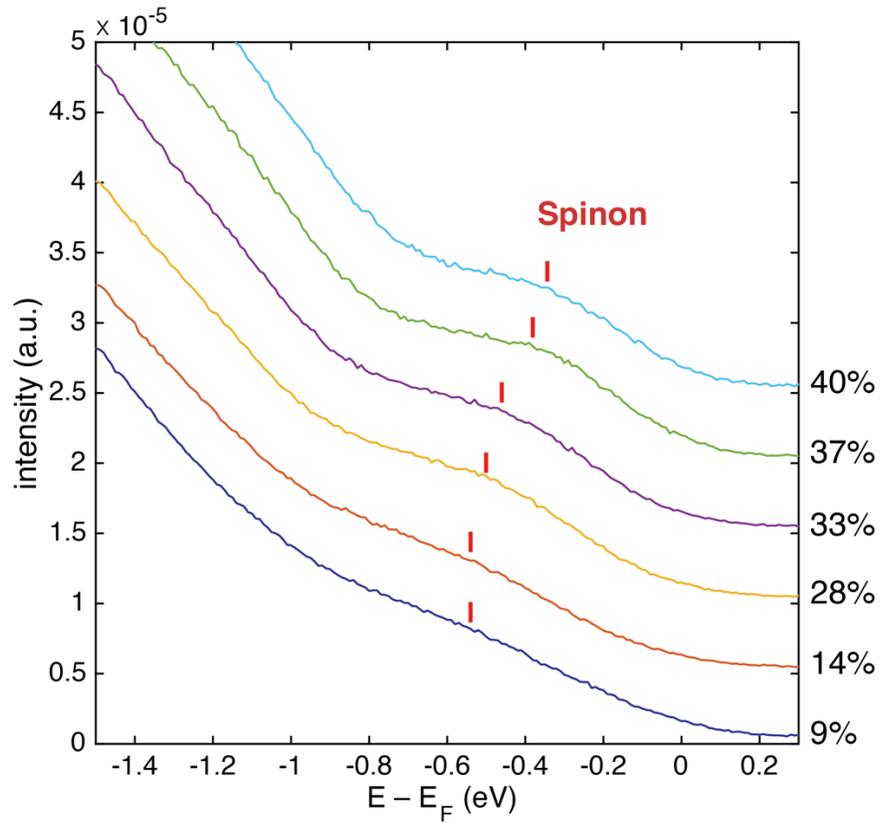

**Fig. S4.**
Doping-dependent energy distribution curves at $k = 0$. Vertical red solid short lines indicate the binding energy fitted using Gaussian peaks, same data as shown in Fig. S2. The spinon peak shifts towards the Fermi level with increasing doping.

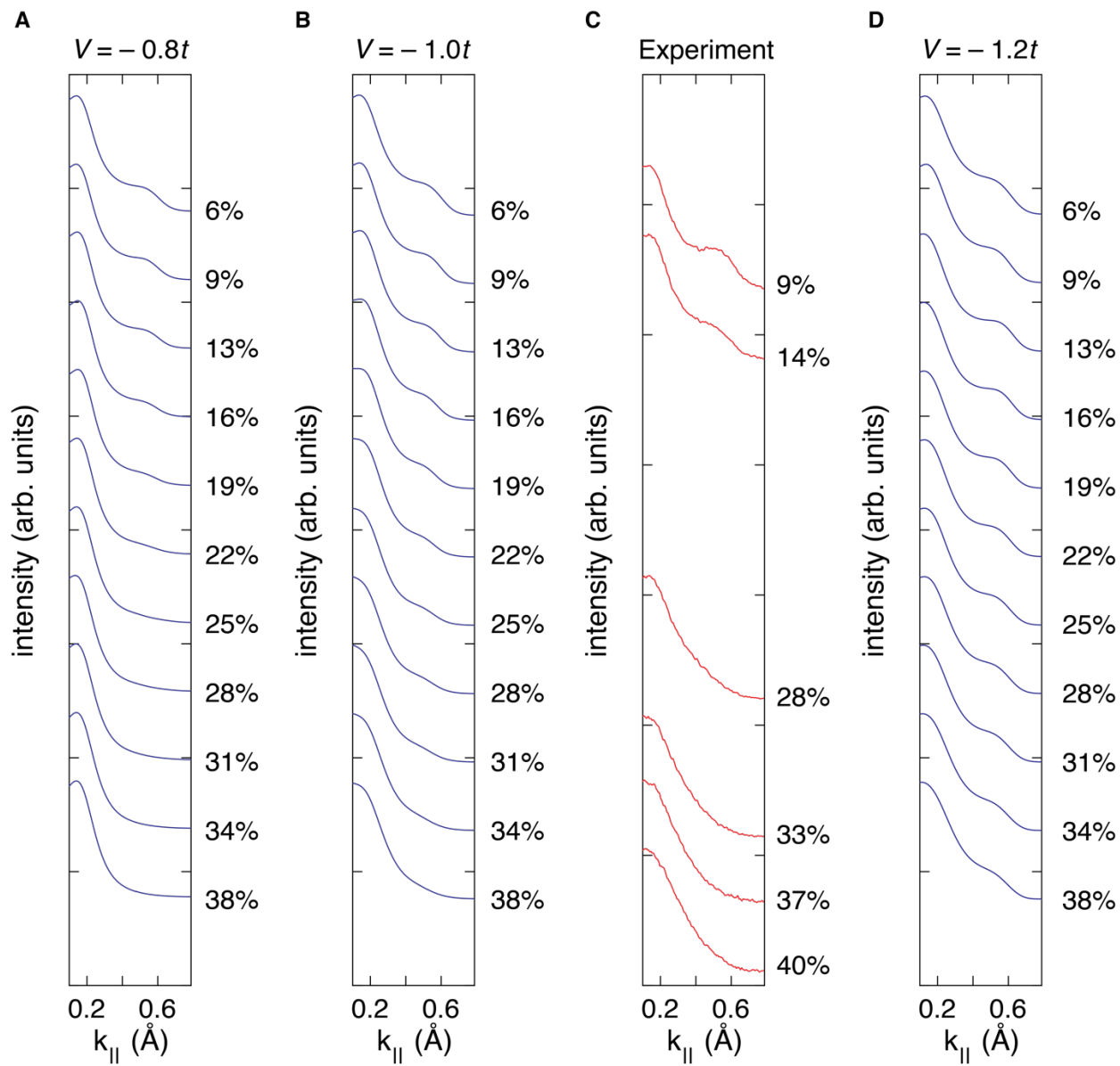

**Fig. S5.**
(A,B,D) CPT simulated momentum distribution curves (MDCs) at different dopings for $V = -0.8t$, $-1.0t$, and $-1.2t$, respectively. The binding energies of these MDCs are chosen so that the major holon peak positions remain similar, for better comparison across different dopings. (C) Experimental MDCs at different dopings. The same set of data as shown in Fig. 4B in the main text.

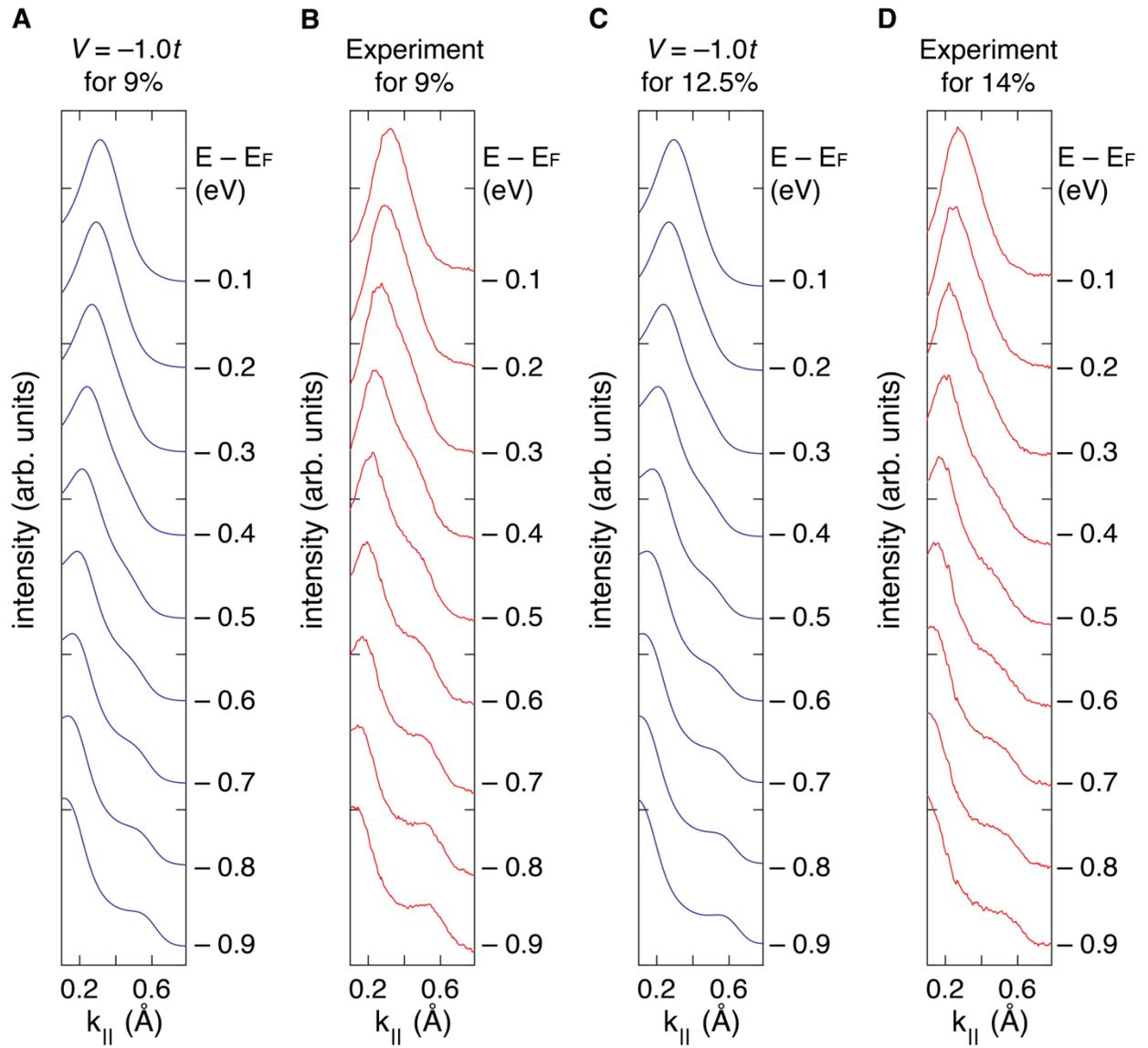

**Fig. S6.**
(A,C) CPT simulated MDCs at different binding energies for $V = -1.0t$, with 9% and 12.5% hole doping, respectively, as compared to experimental conterparts (B,D).

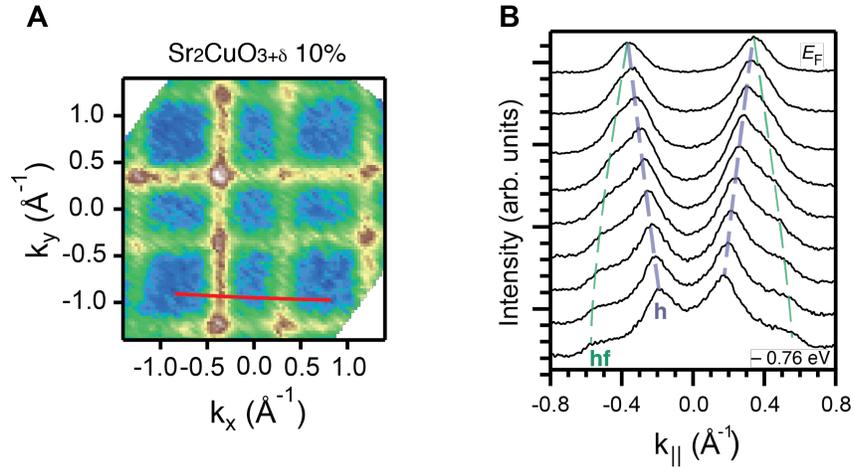

**Fig. S7.**
(A) Fermi surface map of a $Sr_2CuO_{3+\delta}$ thin film with doping of approximately 10%. (B) MDCs at different binding energies along the cut shown in the red curve in (A).